\documentclass[12pt]{article}
\usepackage{graphics}

\newcommand{\eqref}[1]{(\ref{#1})}

\renewcommand{\d}{{\textrm d}}
\newcommand{\ee}{{\rm e}}

\newcommand{\proc}[2]{{\textbf #1}_{#2}}

\newcommand{\xt}{{\proc X t}}

\newcommand{\wt}{{\proc W t}}
\newcommand{\ws}{{\proc W s}}
\newcommand{\ww}{{\bf W}}
\newcommand{\wz}{{\bf W}_0}

\newcommand{\xxt}[1]{{\textbf X}_{t}^{(#1)}}
\newcommand{\intt}[1]{{\int_{0}^{t}#1\,\d s}}
\newcommand{\eq}[1]{{equation (\ref{#1})}}
\newcommand{\EE}[1]{{\textrm I \! \textrm E \left[#1\right]}}
\newcommand{\half}{{\textstyle{\frac12}}}
\newcommand\mean[1]{{\big<#1\big>}}
\newcommand\meanb[1]{{\Big<#1\Big>}}
\newcommand{\eps}{\epsilon}

\def\RR{{\hbox{I\kern-.2em\hbox{R}}}}
\def\PP{{\hbox{I\kern-.2em\hbox{P}}}}
\def\EE{{\hbox{I\kern-.2em\hbox{E}}}}
\def\ZZ{{\hbox{Z\kern-.4em\hbox{Z}}}}
\def\CC{{\hbox{I\kern-.5em\hbox{C}}}}
\newcommand{\reals}{\RR}

\begin{document}

\title{\Huge Stochastic Stokes' drift}

\author{
        Kalvis M. Jansons\(^{1}\) and G.D. Lythe\(^{2}\)
}

\maketitle

\begin{center}
\(^{1}\)Department of Mathematics, University College London,\\
 Gower Street, London WC1E 6BT, England
\end{center}
\begin{center}
\(^{2}\)
Center for Nonlinear Studies,  MS-B258\\
Los Alamos National Laboratory,
NM 87545, USA.\\
\end{center}

\begin{abstract}
\noindent Classical Stokes' drift is the small time-averaged drift 
velocity of suspended non-diffusing particles in a fluid due to the
presence of a wave. We consider the effect of adding diffusion to the
motion of the particles, and show in particular that a non-zero
time-averaged drift velocity exists in general even when the classical
Stokes' drift is zero.  Our results are obtained from a general
procedure for calculating ensemble-averaged Lagrangian mean velocities
for motion that is close to Brownian, and are verified by numerical
simulations in the case of sinusoidal forcing.
\end{abstract}

\smallskip
\centerline{\bf PACS numbers: 02.50-r, 05.40+j, 05.60+w}
\smallskip

A travelling wave in a fluid gives suspended particles a small drift
velocity known as Stokes' drift \cite{stokes,phillips,light}.
  When there is
more than one wave, the drift velocity is calculated by summing the
contributions from each wave \cite{handh,mkg}.  In this letter we
consider the influence of diffusion on the magnitude and direction of
the drift velocity.  As in the classical (diffusionless) case, the
amplitude of the travelling wave is assumed small compared to its
wavelength; a non-zero drift velocity appears at second order in the
amplitude.  In the presence of more than one wave, the classical
Stokes' drift can sum to zero. Diffusion then produces
 a non-zero drift velocity whose magnitude and direction depends on
the diffusivity of the suspended particles.

Several mechanisms for the directed motion of small particles without 
a net macroscopic force have been proposed in the last 10 years 
\cite{buttiker,landauer,magnasco,aandb,doer}.  Interest in such 
`ratchet' effects has been motivated by the search for the 
mechanisms of biological motors, such as the conversion of chemical 
energy into directed motion by protein molecules, and by possible 
applications, such as the separation of particles in solution based on 
their diffusion coefficients.  In both these cases small particles are 
believed to follow dynamics that are overdamped (first derivative in 
time) and noise-dominated.  A drift velocity dependent on the size of 
suspended particles in solution has been produced experimentally using 
an asymmetric periodic potential turned on and off periodically 
\cite{krsap}.  Published theoretical models \cite{mandd,eandd} 
combine a periodic asymmetric potential in one dimension
with non-white fluctuations.

In this letter we consider motion in arbitrary dimensions that is
diffusion-dominated.  There is also a small deterministic forcing
whose amplitude will be used as an expansion parameter; a drift
velocity appears at second order and depends on the diffusivity.  Thus
diffusion due to microscopic motions, for example diffusion of
particles in solution, can be exploited using a carefully-chosen 
combination of forcings to produce a net motion
that depends on the diffusivity.  We illustrate the effect with
sinusoidal forcing and compare our calculations with numerical results
in one and two space dimensions.  It is possible to arrange the wave
motions so that particles of different diffusivities have a
time-averaged drift velocity in different directions, resulting in
what we call `fan-out'.  This may have applications for sorting
particles according to their diffusivities.  We show numerically that
the fan-out can have an angular range of more than 180 degrees.

We first develop an expansion scheme for motion that is overdamped and
diffusion-dominated.  Consider a stochastic process \({\bf
  X}\equiv(\xt)_{t\geq0}\) taking values in \(\reals^m\) and
satisfying the following stochastic differential equation
\cite{gardiner,dynpf}:
\begin{equation}
\d \xt = \epsilon f(\xt,t) \d t + \d \wt, \;\;\; 0\leq\epsilon\ll1.
\label{xtvec}
\end{equation}
The vector \(\xt\) is the particle position at time \(t\).
Its ensemble average, to be denoted below by angled brackets, is
the Lagrangian mean position at time \(t\).
\(\ww\) is an \(m\)-dimensional Brownian motion, with 
\(\wz=0\) and \(\mean{\wt\cdot\wt}=m\sigma^2 t\), i.e.
\(\ww\) represents a purely diffusive motion, with diffusivity 
\begin{equation}
        D = \half \sigma^2.
        \label{D}
\end{equation}
The remaining term in \eqref{xtvec}
is the deterministic forcing, a function
 of Eulerian position \(x\) and time \(t\):
\begin{equation}
f: \reals^m\times\reals^+ \to \reals^m.
\label{fvec}
\end{equation}
The real constant \(\epsilon\) satisfies \(0\le\epsilon\ll 1\).

We now expand in powers of \(\epsilon\). Let
\begin{equation}
        \xt = \xxt0 + \epsilon \xxt1 + \epsilon^{2} \xxt2 + \ldots, \;\;\;
        \mbox{with initial condition} \;\; {\textbf X}_{0} = 0.
        \label{xxt}
\end{equation}
The leading terms of the stochastic equation of motion,
\eq{xxt}, are as follows.
\begin{description}
        \item[\(\epsilon^{0}:\)]
        \begin{equation}
                \d \xxt0 = \d \wt,
        \end{equation}
        giving
        \begin{equation}
                \xxt0 = \wt.
                \label{xxt0}
        \end{equation}

        \item[\(\epsilon^{1}:\)]
        \begin{equation}
                \d \xxt1 = f(\xxt0,t) \d t,
        \end{equation}
        giving
        \begin{equation}
                \xxt1 = \intt{f(\ws,s)}.
                \label{xxt1}
        \end{equation}

        \item[\(\epsilon^{2}:\)]
        \begin{equation}
                \d \xxt2 = (\xxt1\cdot\nabla) f(\xxt0,t)\d t,
        \end{equation}
        giving the second-order drift velocity as
        \begin{equation}
          \frac{\d}{\d t}\xxt2 = 
          \intt{(f(\ws,s)\cdot\nabla) f(\wt,t)},
                \label{xxt2}
        \end{equation}
        where \(\nabla f\) is the spatial gradient of \(f\).
\end{description}

In the classical derivation of Stokes' drift
 there is no motion at zeroth order \cite{stokes,phillips}.
Here the motion at zeroth order is purely diffusive,
with \(\mean{\xxt0}=0\) for any positive time.
When \(f(x,t)\) is a sum of functions that are periodic in 
\(t\) at any fixed \(x\), the drift velocity also vanishes at first order in 
\(\epsilon\).  That is
\begin{equation}
\lim_{t\rightarrow\infty}\frac{1}{t}\mean{\xxt1}=0.
\label{limzm}
\end{equation}
At second-order the drift velocity is in general non-zero. It
is given by the following ensemble average:
\begin{eqnarray}
  U &\equiv & \epsilon^{2}
  \lim_{t\rightarrow\infty}\frac1t\mean{\xt^{(2)}}
  \label{udef}\\
  & = & \epsilon^{2}
  \lim_{T\rightarrow\infty}\left(\frac1T\int_{0}^T
    \meanb{\intt{\left(f(\ws,s)\cdot\nabla\right)f(\wt,t)}}
    \d t\right).
  \label{gendrift}
\end{eqnarray}
In one space dimension the expression \eqref{gendrift} 
reduces to
\begin{equation}
        U \equiv \epsilon^{2}\lim_{T\rightarrow\infty}
        \left(\frac1T\int_{0}^{T}
\meanb{f'(\wt,t)\intt{f(\ws,s)}}\,\d t\right),
        \label{drift}
\end{equation}
where \(f'(x,t)=\frac{\partial}{\partial x}f(x,t)\).         

We now consider the case where the deterministic forcing is a sum of 
sinusoids:
\begin{equation}
f(x,t)=\sum_{\ell=1}^n
A_{\ell} k_{\ell}\cos(k_{\ell}\cdot x -\omega_{\ell} t),
\label{fsinu}
\end{equation}
where \(A_{\ell}\) and \(\omega_{\ell}\) are constants.
The vector \(k_{\ell}\) defines the direction of
propagation of wave \(l\).
The drift velocity \eqref{gendrift} for this case is
\begin{eqnarray}
  U & = & \epsilon^{2}\sum_{\ell=1}^n \Big[
  \displaystyle\lim_{t\to\infty}
  A_{\ell}^2|k_{\ell}|^2k_{\ell}
  \int_0^t\Big<\sin\big(
  - k_{\ell}\cdot(\wt-\ws)+\omega_{\ell} (t-s)\big))\Big>\d s\Big]
  \nonumber \\[10pt]
  & = & \half \epsilon^{2}\sum_{\ell=1}^n \Big[
  \displaystyle\lim_{t\to\infty}
  A_{\ell}^2|k_{\ell}|^2k_{\ell}\int_0^t
  \sin(\omega_{\ell}(t-s))\ee^{-|k_{\ell}|^2D(t-s)}\d s\Big]
  \label{meanux}\\[10pt]
  & = & \half \epsilon^{2}
  \sum_{\ell=1}^n \left[\displaystyle
    A_{\ell}^2|k_{\ell}|^2\frac{k_{\ell}}{\omega_{\ell}}
    \left(1+D^2\frac{|k_{\ell}|^4}{\omega_{\ell}^2}\right)^{-1}\right].
  \nonumber
\end{eqnarray}
Each wave makes a contribution to the drift velocity in its direction 
of propagation.  For \(D=0\), the weighting factor is proportional to 
the square of the amplitude.  This is the classical result obtained by 
a transformation from Eulerian to Lagrangian coordinates 
\cite{stokes,phillips}.  In the case of a surface wave over deep 
water, the first order motion of a suspended particle is a circle with 
radius \(A_{\ell}k_{\ell}/\omega_{\ell}\); the quantity 
\(A_{\ell}^2|k_{\ell}|^2{k_{\ell}}/{\omega_{\ell}}\) is proportional 
to the time-averaged momentum per unit area \cite{phillips,light}.  In 
the presence of diffusion, the contribution from wave \(\ell\) is 
reduced by the dimensionless factor \(\left(1+ 
\alpha_{\ell}^2\right)^{-1}\), where \(\alpha_{\ell} = 
D{|k_{\ell}|^2}{\omega_{\ell}^{-1}}\).  Diffusion reduces the Stokes 
drift due to any one wave by smearing out the distribution of 
particles, working against the tendency of particles to spend longer 
in regions where the force acts in the direction of propagation than 
in those where the force acts in the opposite direction.  The 
attenuation is strongest for waves with large wavenumbers or small 
velocities.

Dependence of drift velocity on diffusion can be exploited as follows:
there is in general a non-zero drift velocity due to diffusion even
when the classical Stokes' drift is zero.  We write the drift velocity
\eqref{meanux} as a sum of the classical Stokes' drift and a
diffusion-dependent contribution: 
\begin{equation} 
U=U_0 + U_{\rm s},
\label{decomp} 
\end{equation}
 where 
\begin{equation} U_0=U|_{D=0}=\half \epsilon^{2} \sum_{\ell=1}^n
      \displaystyle
        A_{\ell}^2|k_{\ell}|^2\frac{k_{\ell}}{\omega_{\ell}}
\label{uzdef}
\end{equation}
and
\begin{equation}
U_{\rm s}
= - \half \epsilon^{2}
\sum_{\ell=1}^n \left[
      \displaystyle
        A_{\ell}^2|k_{\ell}|^2\frac{k_{\ell}}{\omega_{\ell}}
        \frac{\alpha_{\ell}^2}{1+\alpha_{\ell}^2}
\right].
\label{uuz}
\end{equation}

The classical Stokes drift \(U_0\) can be made to vanish
by choosing a forcing \(f(x,t)\) consisting of two wave trains
propagating in opposite directions. For the latter  example, 
we can work in one space dimension, defined as the
direction of propagation of wave \(\ell = 1\):
\begin{equation}
        f(x,t) = A_{1} k_1\cos(k_{1}x-\omega_{1}t+\phi_{1})
        + A_{2} k_2\cos(k_{2}x-\omega_{2}t+\phi_{2}),
        \label{eg1}
\end{equation}
where \(A_{i}\), \(k_{i}\), \(\omega_{i}\) and \(\phi_{i}\) 
(\(i=1,2\)) are constants and \(k_1k_2<0\).  For simplicity, we 
suppose that \(k_{1}\neq \pm k_{2}\) and \(\omega_{1} \neq \pm 
\omega_{2}\); this avoids cross-terms in the classical Stokes' drift.  
The drift velocity including diffusion is then given by
\begin{equation}
U      = \half \epsilon^{2}
      \displaystyle
        \left(
        A_1^2\frac{k_{1}^3}{\omega_{1}}
        \left(1+\alpha_1^2\right)^{-1}
        + A_2^2\frac{k_{2}^3}{\omega_{2}}
        \left(1+\alpha_2^2\right)^{-1}
        \right).
\end{equation}
To set \(U_{0} = 0\) requires
\(
        {A_{1}^{2}k_{1}^3}/{\omega_{1}}
        = -{A_{2}^{2}k_{2}^3}/{\omega_{2}}.
\)
Then \(U=U_{\rm s}\) where
\begin{equation}
        U_{\rm s} = \half \epsilon^{2}A_1^2        
        \frac{k_{1}^3}{\omega_{1}}
        \left(
        \left(1+\alpha_1^2\right)^{-1}
        - \,
        \left(1+\alpha_2^2\right)^{-1}
        \right).
        \label{Ddrift2}
\end{equation}
For large diffusivity the drift velocity tends to zero 
because the contribution of each wave tends to zero.
Thus there is an intermediate value of diffusivity
that maximises \(U_{\rm s}\).
If the forcing frequencies and wavenumbers are fixed
and \(U_0=0\), the drift attains its maximum
at the value of \(D\) satisfying \(\alpha_1\alpha_2=1\).

Figure \ref{driftfig} shows the drift velocity as a function of
diffusivity with the forcing a sum of two sinusoids for a choice of
parameters that gives \(U_0=0\). In Figure \ref{trajfig} the
calculated drift is compared with numerical results, with the same
choice of parameters and \(D=0.125\).  The solid line in Figure
\ref{trajfig}(a) is the mean value of \(\xt\) as a function of time,
averaged over 10000 numerical realizations of the stochastic
differential equation \eqref{xtvec}, and the dotted line is \(Ut\),
with \(U\) given by \protect\eqref{Ddrift2}.  In Figure
\ref{trajfig}(b) we show, as a function of time, the difference
between the numerically-calculated mean displacement and \(Ut\).
Figure \ref{trajfig}(c) demonstrates that the motion is close to
Brownian; a histogram of values of \(\xt\) at \(t=1000\), \(R(y)\), is
compared with the Gaussian probability density function with mean \(
Ut \) and variance \(\sigma^2 t\) (solid line).

In general the expressions \eqref{meanux} and
\eqref{uuz} are vector relations.  Thus,
in more than one space dimension, the direction as well as the
magnitude of the drift velocity depends on the diffusivity, producing
fan-out of the drift velocity vectors.  We illustrate this effect in
Figure \ref{fanfig}, constructed with the forcing being a sum of four
sinusoids in two dimensions.  In (a), the vector \(A_{\ell}k_{\ell}\)
is shown for each of the four waves.  The parameters are
\protect\(A_1=1.0\), \protect\(A_2=0.8\), \protect\(A_3=0.7\),
\protect\(A_4=0.7\); \protect\(k_1=(1.0,0.0)\),
\protect\(k_2=(2.0,-4.0)\), \protect\(k_3=(-3.0,0.7)\),
\protect\(k_4=(-0.96,4.56)\).  We take \(\omega=v k\) with \(v=1\).
Figure \ref{fanfig}(b) depicts the fan-out in the directions and
magnitudes of the drift velocities for nine different values of
diffusivity.  Each arrow is \(U_{\rm s}\) for one of the following values of
\(D\): \(D=0.1\) (leftmost arrow), \( 0.2 \ldots 0.9\) (rightmost
arrow). For larger values of \(D\), the direction of \(U\) approaches
more closely that of \(k_1\).

The fan-out effect shown in Figure
 \ref{fanfig} is due to the different rates at which
the contribution from waves decreases as the diffusivity
is increased, destroying the exact cancellation imposed
at \(D=0\). More light is shed by considering
 the small-diffusivity and large-diffusivity  limits of \eqref{uuz}.
\begin{enumerate}
        \item\label{caseone} If
\(
D{|k_{\ell}|^2}/{\omega_{\ell}}\ll 1 \quad \forall l
\)
then
\begin{equation}
U_{\rm s} 
= - \half \epsilon^{2}D^2
\sum_{\ell=1}^n \left[
      \displaystyle
        A_{\ell}^2\frac{|k_{\ell}|^6}{\omega_{\ell}^3}k_{\ell}
+\ldots\right].
\label{usmalld}
\end{equation}

        \item\label{casetwo} If
\(
D{|k_{\ell}|^2}/{\omega_{\ell}}\gg 1 \quad \forall l
\)
then
\begin{equation}
U_{\rm s} 
=  \half \frac{\epsilon^{2}}{D^2}
\sum_{\ell=1}^n \left[
      \displaystyle
        A_{\ell}^2\frac{\omega_{\ell}}{|k_{\ell}|^2}k_{\ell}
+\ldots\right].
\label{ubigd}
\end{equation}
\end{enumerate}
In the limit of small diffusivity (\ref{caseone}) the drift velocity
is proportional to \(D^2\) and the direction is approximately opposite
to that of the wave with the largest value of
\(A^2{|k|^6}{\omega^{-3}}\).  In the opposite limit (\ref{casetwo})
the drift velocity is proportional to \(D^{-2}\) and the direction is
approximately parallel to that of the wave with the largest value of
\(A^2{\omega}{|k|^{-2}}.\)

In summary, we derive a general expression for the drift velocity of
diffusing particles from a stochastic asymptotic expansion scheme for
motion that is Brownian plus a small deterministic forcing.  The drift
velocity is in general non-zero even when the classical Stokes' drift
vanishes.  For example, several counterpropagating sinusoidal forcings
produce a drift velocity that depends on the diffusion coefficient and
the intensities, frequencies and wavenumbers of the forcings.  Thus,
given a collection of particles with different diffusivities, the
deterministic forcings can be tuned to separate particles of a
particular type by optimizing their stochastic Stokes' drift.

\clearpage

\clearpage

\begin{figure}
\centering
\includegraphics{stokdrifteps.aux}
\caption{
{\it Stochastic Stokes' drift in one dimension}.
There is a non-zero drift velocity due to diffusivity even though
 the classical Stokes' drift vanishes, due to the different rates at
which the contributions from each wave decrease as the diffusivity is
increased.  The drift velocity, \protect\eqref{Ddrift2}, is given as a
function of diffusivity for \(\eps=0.1\), \( A_1=k_1=\omega_1=1 \),
\(k_2=-2.42 \),
 \(\omega_2=0.47 \).
}
\label{driftfig}
\end{figure}
\begin{figure}
\centering
\includegraphics{drifteps.aux}
\caption{
{\it Stochastic Stokes' drift:
comparison with numerical solution for
sinusoidal forcing in one dimension}.
(a) Mean value of \(\xt\).
(b) Difference between the numerical mean value
and the second-order result \protect\eqref{Ddrift2}.
(c) Distribution of \(\xt\) at \(t=1000\),
\(\sigma=0.5\).
}
\label{trajfig}
\end{figure}
\begin{figure}
\centering
\includegraphics{fan4eps.aux}
\caption{ 
{\it Stochastic Stokes' drift in two dimensions.} 
(a) Directions and magnitudes of the four sinusoidal forcings.  The 
classical Stokes' drift \protect\eqref{uzdef} is zero.  (b) Resulting 
stochastic Stokes' drift as a function of diffusivity.  Here 
\protect\(D=0.1,\ldots,0.9\) (largest \protect\(D\) on the right).  In 
the axis labels, the subscripts indicate vector components.  Note the 
fan-out of more than 180 degrees.  }
\label{fanfig}
\end{figure}

\end{document}